\documentstyle[12pt]{article}

\newcommand{\be}{\begin{equation}}
\newcommand{\ee}{\end{equation}}
\newcommand{\ba}{\begin{eqnarray}}
\newcommand{\ea}{\end{eqnarray}}

\newcommand{\no}{\nonumber}

\begin{document}

\begin{flushright}
March, 1998 \\
\end{flushright}

\begin{center}
{\Large \bf S-duality and Strong Coupling Behavior of Large $N$ 
Gauge Theories with ${\cal N}=4$ Supersymmetry}
\end{center}

\vspace{5mm}

\begin{center}
Tohru Eguchi

\bigskip

{\it Institute for Theoretical Physics,

\medskip

University of California,

\medskip

Santa Barbara,

\medskip

CA 93106-4030, U.S.A.

\medskip

and 

\medskip

Department of Physics, Faculty of Science, 

\medskip

University of Tokyo

\medskip

Tokyo 113, Japan}
\end{center}

\bigskip

\bigskip

\begin{abstract}
We analyze the strong coupling behavior of the large $N$ 
gauge theories in 4-dimensions with ${\cal N}=4$ supersymmetry 
by making use of S-duality.  We show that at large values of 
the coupling constant $\lambda=g_{YM}^2N$ 
the $j$-th non-planar amplitude $f_j(\lambda) \hskip1mm (j=0,1,2 \dots)$ 
behaves as $f_j(\lambda)\approx \lambda^j$.  Implication of
this behavior is discussed in connection with the supergravity 
theory on $AdS_5\times S^5$ suggested by the CFT/AdS correspondence. 
S-duality of the gauge theory corresponds to the duality between the 
closed and open string loop expansions in the gravity/string theory.

\end{abstract}

\newpage 
Understanding the dynamics of $SU(N)$ gauge theories in 4 dimensions 
in the large $N$ limit has been an outstanding problem during the last 
few decades. 
It was first pointed out by `t Hooft \cite{tH1}
that when one fixes the value of the 
coupling constant $\lambda=g_{YM}^2N$, 
Feynman diagrams 
of perturbation theory organize themselves into  
Riemann surfaces and the amplitudes corresponding to higher 
genus Riemann surfaces are suppressed by powers of $1/N$. Thus the planar
diagrams dominate in the large $N$ limit.
In ref.\cite{tH2}, the convergent nature of the planar
perturbation series for each fixed value of the genus has been established
(modulo the problem of renormalization).
If we denote as $f_j(\lambda)$ the genus $j$ amplitude of some physical 
quantity, the weak coupling perturbative series $f_j(\lambda)=\sum_{\ell=0}
a_{j,\ell}\lambda^{\ell}$ converges in a finite range $|\lambda|<\lambda_c$ 
(the radius of convergence $\lambda_c$ is believed to be universal for all 
$j$).  $f_j(\lambda)$ will be 
regular on the positive real axis $\lambda>0$ while it 
has a singularity at $-\lambda_c$ on the negative real axis. 

We expect the functions $f_j(\lambda) \hskip1mm (j=0,1,2\cdots)$ to have
analytic continuations into the strong coupling region $|\lambda|
>\lambda_c$. In this regime they will possess convergent  $1/\lambda$ 
expansions of the form $f_j(\lambda)=\sum_{k=0} b_{j,k}\lambda^{a-bk}\hskip1mm
(b>0)$.

Planar perturbation theory has been considered exclusively in the 
weak coupling regime $|\lambda|<\lambda_c$ so far.
Recently, however, interesting observations have been made on the relation 
between the supergravity/string theories in the space-time and the 
field theories defined on branes on its boundary \cite{K,GKT,GK,Mal1} 
(CFT/AdS correspondence). 
Under this correspondence the 4 dimensional 
${\cal N}=4$ supersymmetric Yang-Mills theory 
is related to a supergravity theory on 
$AdS_5\times S^5$. If one denotes the radius of the space 
$AdS_5/S^5$ as $R$ and 
the string coupling and length as $g_{st}$ and $\ell_s$, respectively,
the mapping is given by \cite{Mal1}
\be
{R \over \ell_s}=(4\pi\lambda)^{1/4}, \hskip15mm g_{st}=g_{YM}^2.
\label{map}\ee
Thus the strong coupling region of gauge theory is mapped to the
semi-classical regime of gravity where the size of the space-time is
much larger than the string length $\ell_s$. (On the other hand 
the weak coupling regime is mapped
to a region of gravity/string theory where the space-time is much 
smaller than $\ell_s$.) In this context of particular interest is the 
possibility of a strongly
coupled gauge system being described by a classical gravity theory in  
some particular background. In a recent work \cite{W2} 
the thermodynamic properties of ${\cal N}=4$
Yang-Mills theories are derived using classical solutions in
$AdS$ space.

In order to have a better understanding of the CFT/AdS correspondence  
it is important to gain further insights to the strong 
coupling dynamics of large $N$ gauge theories.
In this paper we would like to analyze
the strong coupling 
behavior of the large $N$ gauge theory by making use of its S-duality.
In the following we first consider a functional
equation for the dependence of the coupling constant $\lambda=g_{YM}^2N$ 
on some physical amplitude derived by imposing the invariance 
under S-duality $g_{YM} \leftrightarrow 1/g_{YM}$. 
It is easy to find a consistent solution to the functional
equation where the $j$-th non-planar amplitude $f_j(\lambda)$ 
has a behavior $f_j(\lambda)\approx \lambda^j$ for large $\lambda$.
We then discuss our result in connection with the expected behavior of the
supergravity/string theory amplitudes in $AdS_5/S^5$ space.

According to our analysis the planar amplitude behaves as $f_0(\lambda) 
\approx const$  as $\lambda \rightarrow \infty$. This implies that
the perturbative D-brane computation of thermodynamics of near extremal
black hole at small $\lambda=g_{st}N$ may be smoothly extrapolated to the 
large-$\lambda$ region of classical gravity. In particular
the discrepancy of a factor 3/4 in the computation of the entropy
of near extremal black 3-brane in \cite{GKP,KT} may be accounted for
if $f_0(\lambda=\infty)=3/4$. Unfortunately, the S-duality does not have enough
power to determine the functional form of $f_0(\lambda)$ explicitly.

Let us consider the large $N$ expansion of some physical amplitude
in ${\cal N}=4$ Yang-Mills theory,
\be
F(\lambda,N)=\sum_{j=0}f_j(\lambda)N^{2-2j},  \hskip5mm \lambda=g_{YM}^2N.
\label{amp}\ee
$f_0(\lambda)$ denotes the sum over amplitudes of planar Feynman 
diagrams. Its weak coupling expansion, 
$f_0(\lambda)=\sum_{\ell=0}a_{\ell}\lambda^{\ell}$ is
expected to have a finite radius of convergence $\lambda_c$ 
around the origin in the 
complex $\lambda$ plane \cite{tH2}. 
Similarly $f_j(\lambda) (j\ge 1)$ gives a sum over the amplitudes of diagrams 
with $j$ handles. As is well-known these amplitudes are suppressed by 
$N^{-2j}$ in the $1/N$
expansion. Weak coupling expansion of $f_j$ has the same 
radius of convergence $\lambda_c$.

Let us assume that the above amplitude is invariant under the $S$-duality 
transformation
$g_{YM} \rightarrow 1/g_{YM}$. In terms of $\lambda$ the transformation is 
given by $\lambda\rightarrow N^2/\lambda$. Hence
\be
\sum_{j=0}f_j(\lambda)N^{2-2j}=\sum_{j=0}f_j(N^2/\lambda)N^{2-2j}.
\label{damp}\ee
(We may consider a more general case when the amplitude is a modular form
with a certain weight. It is straightforward to modify the following 
discussions for these cases.) 

Now let us determine the large $\lambda$ behavior of functions $f_j(\lambda)$. 
We assume that the functions $f_j(\lambda)$ can be analytically continued
to $|\lambda|>\lambda_c$ and has a convergent expansion in $1/\lambda$.
By comparing the both sides of the functional relation (\ref{damp}) we find 
that the following {\it Ansatz}
\be
f_j(\lambda)=\sum_{\ell=0}C_{j,\ell}\lambda^{j-\ell} \hskip24mm 
\label{fjst}\ee
is consistent. 
By substituting (\ref{fjst}) 
into the RHS of (\ref{damp}) we obtain
\be
\sum_{j,\ell}C_{j,l}\left( {N^2 \over \lambda} \right)^{j-\ell}N^{2-2j}
=\sum_{j,\ell}C_{j,\ell}\lambda^{\ell-j}N^{2-2\ell}.
\ee
On the other hand the LHS becomes
\be
\sum_{j,\ell}C_{j,\ell}\lambda^{j-\ell}N^{2-2j}.
\ee
Hence the {\it Ansatz} is consistent if the coefficients 
$C_{j,\ell}$ are symmetric
\be
C_{j,\ell}=C_{\ell,j}.
\label{symm}\ee
In terms of the original gauge coupling $g_{YM}$ the amplitude $F$
is written as
\be
F(g_{YM},N)=\sum_{j,\ell}C_{j,\ell}\lambda^{\ell-j}N^{2-2\ell}=
\sum_{j,\ell}C_{j,\ell}{{g_{YM}}^{\ell}{g_{YM}}^{-j} \over N^{\ell+j-2}}
\ee
which is manifestly symmetric under $g_{YM}\leftrightarrow 1/g_{YM}$ due to
the symmetry of the expansion coefficients eq.(\ref{symm}).

Symmetry of the coefficients $C_{j,\ell}=C_{\ell,j}$ also implies 
that the 
expansion coefficients of the planar amplitude
\be
f_0(\lambda)=\sum_{\ell=0}C_{0,j}\lambda^{-j}
\ee
determine the leading coefficients of the non-planar amplitudes
$f_j(\lambda) \approx C_{0,j}\lambda^{j}$.

In the above we have considered the strong coupling region 
of the large $N$ 
gauge theory where quantum Yang-Mills theory is 
expected to be "dual"
to the semi-classical supergravity/string theory on $AdS_5\times S^5$.
The precise mapping of the parameters between these theories is given by 
\cite{Mal1}
\be
{R \over \ell_s}=(4\pi\lambda)^{1/4}, \hskip5mm {R \over \ell_{P,10}}=
(4\pi N)^{1/4}
\label{map2}\ee
where $\ell_{P,10}=g_{st}^{1/4}\ell_s$ is the 10-dimensional Planck
length. Thus the large $N$ limit corresponds to the limit of small curvature 
of the $AdS_5/S^5$ space and the large $\lambda$ corresponds to the 
suppression of string theoretic corrections.

In string theory the structure of the perturbative expansion in the 
presence 
of $N$ coincident Dirichlet branes is given by
\be
F(g_{st},N)=\sum_{j,b}D_{j,b}(g_{st}N)^b {g_{st}}^{2j-2}
\ee
where $b$ counts the number of open-string loops attached to the D-branes
and $j$ counts the number of closed-string loops. $D_{j,b}$'s are some
numerical constants. Here the factor $N$ 
appears due to Chan-Paton factors which distinguish different D-branes.
Using $g_{st}N=\lambda$ one may rewrite the above expansion as
\be
F(\lambda,N)=\sum_{j,b}D_{j,b}\lambda^{b+2j-2}N^{2-2j}.
\ee
This exactly reproduces the weak coupling planar perturbation series in 
gauge theory.
Note that in this formula the double-lines representing a gluon
in Yang-Mills theory are regarded as describing the motion of end 
points of an open string and hence $b$ equals the number of index-loops
in the standard large $N$ counting. The sum
over open-string loops converges in the weak coupling region
$|\lambda|<\lambda_c$. According to our assumption 
the series may be continued to the strong
coupling region and yields a $1/\lambda$ expansion,
\be
F(\lambda,N)=\sum_{j,\ell}C_{j,\ell}\lambda^{j-\ell}N^{2-2j}.
\ee
Thus the S-duality symmetry $C_{j,\ell}=C_{\ell,j}$ of gauge theory
corresponds to a dual
relationship between the open-string and closed-string loop expansions
in gravity/string theory.

It is well-known that the region of validity of the D-brane perturbation theory
in the study of thermodynamic properties of black holes is 
where the size of the black hole is much smaller than the string length
(for a review of D-brane approach to black hole physics, see, for instance 
\cite{Mal2}).
This corresponds to the weak coupling regime $|\lambda|<\lambda_c$ in the
present context. On the other hand in the region of validity of classical 
gravity the size of black hole must be much larger than the string 
length and hence $|\lambda|>\lambda_c$. The crossover 
range $\lambda \approx \lambda_c$ corresponds to
the matching point in the analysis of \cite{HP}.
In the case of the computation of the entropy of extreme black holes 
\cite{SV,CM} 
predictions of the D-brane perturbation theory can be smoothly
continued to the large coupling region due to the topological nature
of BPS states. On the other hand in the case of black holes away from
the extremality the stability of the extrapolation is not guaranteed 
\cite{Das}.
Our result $f_0(\lambda) \approx const$ as $\lambda \approx \infty$,
however, seems to assure a smooth dependence on the coupling constant
$\lambda$ as far as the
planar amplitudes are concerned. 
In particular the values of $f_0$ at $\lambda=0$ and $\infty$
could differ only by some numerical factor. The discrepancy by a factor 3/4
in the computation of entropy of near extremal black three-brane 
\cite{GKP,KT} may in fact be explained in terms of strong coupling effects
of gauge theory.

According to recent papers \cite{KS,LNV,BKV,Ka,BJ}
planar amplitudes of 4-dimensional gauge 
theories with vanishing beta functions
are identical irrespective of the numbers of supersymmetries
${\cal N}=4,2,1$. Hence our results $f_0(\lambda)\approx const$
at $\lambda \approx \infty$ would also hold in a large class of 
Yang-Mills theories
with ${\cal N}=2,1$ supersymmetries. It will be extremely interesting 
if we could test this prediction.

We now rewrite the gauge theory amplitude $F$ using the relation
(\ref{map2}),
\ba
&&F=\sum_{j,\ell=0}C_{j,\ell}\lambda^{j-\ell}N^{2-2j}
=N^2\sum_{j,\ell=0}C_{j,\ell}\lambda^{-\ell}
\left(g_{st}^2 \over \lambda \right)^j \no \\
&&=N^2\sum_{j,\ell=0}C_{j,\ell}
\left(4\pi \left(\ell_s \over R\right)^4\right)^{\ell}
\left(4\pi g_{st}^2 \left({\ell_s \over R}\right)^4\right)^j. 
\ea
If one uses the  10-dimensional Newton constant $G_N$
\be
16\pi G_N\equiv2\kappa^2
=(2\pi)^7g_{st}^2{\ell_s}^8,
\label{new}\ee 
F is further rewritten as
\ba
&&F={\pi^4R^8 \over 2G_N}\sum_{j,\ell=0}C_{j,\ell}
\left(4\pi \left(\ell_s \over R\right)^4\right)^{\ell}
\left({8 \over (2\pi)^6} {G_N \over R^4{\ell_s}^4}\right)^j.
\label{sugra}\ea
The overall factor $R^8/G_N$ (proportional to 
$N^2$) of (\ref{sugra})
has a natural form from the point of view of classical gravity
since $R^8/G_N$ is the unique dimensionless combination 
of the 10-dimensional Newton constant and the radius $R$ of space-time.
Open-string couplings to the D-branes are expanded in power series of 
$(\ell_s/R)^4$ and generate $\alpha'$ corrections in (\ref{sugra}). 
The sum over the closed-string 
loops, on the 
other hand, are expanded in power series of $G_N/R^4{\ell_s}^4$. This factor
is enhanced from the tree level classical value $G_N/R^8$
by $R^4/{\ell_s}^4$. This enhancement may be interpreted 
as being due to 
ultra-violet divergences of classical gravity at quantum level
being cured by string theory 
with an effective cut-off at $\ell_s$.

\bigskip

I would like to thank ITP, Santa Barbara for its
hospitality where this work was done. I also 
would like to thank T.Banks, H.Ooguri,
S.Shenker and T.Yoneya for discussions. This research was supported in part by
the National Science Foundation under Grant No. PHY94-07194.

\end{document}